# An Empirical Study on the Amount of Changes Required for Merge Request Acceptance

Samah Kansab*, Mohammed Sayagh*, Francis Bordeleau*, Ali Tizghadam†
*Software and IT Engineering Department,
École de technologie supérieure (ÉTS), Montreal, Canada
{samah.kansab.1, mohammed.sayagh, francis.bordeleau}@etsmtl.net
†TELUS Communications, Toronto, Canada
ali.tizghadam@telus.com

***Abstract*—Code review (CR) is essential to software development, helping ensure that new code is properly integrated. However, the CR process often involves significant effort, including code adjustments, responses to reviewers, and continued implementation. While past studies have examined CR delays and iteration counts, few have investigated the effort based on the volume of code changes required—especially in the context of GitLab's Merge Request (MR) mechanism, which remains underexplored. In this paper, we define and measure CR effort as the amount of code modified after submission, using a dataset of over 23.6k MRs from four GitLab projects. We find that up to 71% of MRs require adjustments post-submission, and 28% of these involve changes to over 200 lines of code. Surprisingly, this effort is not correlated with review time or the number of participants. To better understand and predict CR effort, we train an interpretable machine learning model using metrics across multiple dimensions: text features, code complexity, developer experience, review history, and branching. Our model achieves strong performance (AUC 0.84–0.88) and reveals that complexity, experience, and text features are key predictors. Historical project characteristics also influence current review effort. Our findings highlight the feasibility of using ML to explain and anticipate the effort needed to integrate code changes during review.**

## I. INTRODUCTION

Code review is a critical part of the software development lifecycle, ensuring code quality, reliability, and maintainability [32]. It involves examining code contributions before integration into the main codebase and can be conducted through various mechanisms, such as informal inspections, pair programming, or tool-supported platforms like pull requests (GitHub) or merge requests (GitLab). In this study, we focus on Merge Request (MR), GitLab's structured mechanism for managing code reviews[1]. Prior research highlights the benefits of code review, including improved readability [1], reduced defect rates [21], and enhanced team knowledge sharing. Beyond detecting bugs, code reviews foster collaboration and promote the exchange of expertise among team members.

In recent years, the need for quantitative metrics to evaluate and improve the effectiveness of the code review process has been increasingly recognized by researchers and practitioners. Notable contributions to the literature have explained various metrics, such as the time to first comments [10], the characteristics of code patches that attract more reviewers [31], the intentions behind code changes [33], etc. To estimate the software effort that developers invest to complete the code review process, researchers have relied on time as a metric to measure the effort expended during code review [4, 35, 20, 13]. They also studied other metrics that can impact time and increase efforts, such as tools to identify the code reviews with more priority [7, 25], the early prediction for merged or rejected code reviews [14], etc.

However, relying solely on time can be insufficient to measure code review efforts because developers may be working, updating, and reviewing code simultaneously within the same amount of time. This becomes particularly noticeable when faced with tight deadlines or urgent tasks, making it difficult to accurately assess the effort spent on the code review process based only on time. For example, we found an MR[2] that changed two lines of code and took more than three months to be integrated, while an MR[3] with 4,370 changes took only four hours to be merged.

In this paper, we empirically investigate review effort by measuring the amount of changes an MR undergoes between its creation and acceptance. These changes typically involve addressing reviewer feedback or modifying the submitted code on GitLab. We define the amount of changes required to accept a MR *as the sum of added and removed lines in all commits made after the MR is created. This reflects the actual volume of correction or adjustment needed before integration.* The closest work to ours includes the studies by Wang et al. [34] and Wang, Bansal, and Nagappan [33], which measured effort using the number of iterations (patchsets) submitted before merging. However, the number of iterations does not necessarily reflect the amount of changes—a single patchset may involve a few lines, while another may modify hundreds.

Our empirical study is guided by the following research questions:

- **RQ1:** ***How much changes are required to merge a MR?***

[1]https://docs.gitlab.com/ee/user/project/merge _requests/

[2]https://gitlab.com/gitlab-org/gitlab-runner/-/merge_requests/1780
[3]https://gitlab.com/gitlab-org/gitlab-runner/-/merge_requests/520

We found that up to 73% of the MRs exhibit updating at least one line of code during the code process. 28% of the MRs require changing (adding and/or removing) at least 200 lines of code before being merged. Such an effort is not correlated with the number of involved practitioners in the review process, nor the time to merge a change which was already explained in the literature. **This finding underscores the need to understand the factors that influence the amount of changes a MR has to exhibit before being accepted**.

- **RQ2:** ***How effective are machine learning models at classifying the amount of changes in MRs?*** In this RQ, we compare different binary machine learning models to classify MRs according to the efforts to accept them (”large” and ”small”). Our results revealed that the random forest model is the best classifier, with a median AUC that ranges between 0.84 and 0.88. All of our feature dimensions (i.e., complexity, text, collaboration, history, branching and experience) are relevant in the classification of MRs based on their amount of changes to be accepted.
- **RQ3:** ***What are the most important features that influence the amount of changes required to accept a MR?*** In this RQ, we studied the importance and the impact of each feature on the efforts to accept a MR. We find that Complexity-related features consistently stand out as the primary two factors impacting amount of changes for accepting MRs across all projects. Experience and text-related features also hold significance, consistently ranking within the top five in terms of importance. Additionally, it is noteworthy that, in some cases, historical features exert an impact on ongoing MRs, leading to an increase in the amount of changes for accepting MRs.

This study underscores the importance of effectively managing efforts to accept a MR and highlights the need for a comprehensive understanding of this metric. Through the application of machine learning, we present a classification model that is able to categorize MRs based on the amount of efforts to be accepted, thereby identifying instances that require significant adjustments prior to acceptance. In addition, our analysis provides practitioners with valuable insights into the primary factors influencing such efforts and their subsequent impact. The same model can be also used for predicting the efforts required for a MR at the time of the MR creation. With this understanding, practitioners can adopt informed strategies to mitigate MRs that require a large amount of efforts to be accepted, and manager better efforts estimation to organize their resources, thereby promoting efficiency and code quality in software development efforts.

Our replication package, which includes both the collected data and analysis scripts, is publicly available at: https://figshare.com/s/632a15054ffb881ae72d

## II. Background and Related Work

Code review is a peer-based process to ensure submitted code meets quality standards before integration into the main codebase [32]. While platforms like Gerrit and GitHub support code review, this study focuses on GitLab’s Merge Request (MR) mechanism. In GitLab, an MR involves proposing changes from a dedicated branch, running local checks, and requesting review for integration into a target branch. MRs can be ”Opened,” ”Closed” (if not needed), or ”Merged” (when accepted). Contributors—including authors, reviewers, commenters, and integrators—collaborate throughout the process to assess and improve code quality.

Several studies [4, 10, 31, 20, 14] investigated the code review duration, trying to optimize it. For instance, Chouchen et al. [4] developed a machine learning model to predict the time to complete the code review process and explain the factors that impact this metric. Hasan et al. [10] found that a shorter time-to-universal-first-response (bot or human) is generally linked to a shorter pull request lifetime, but this relationship may not be accurate in bot-first pull requests due to developers’ limited involvement in the review process. Bosu and Carver [3] found that major developers receive a quicker initial feedback on their review requests, leading to a shorter overall review time. Jiang, Adams, and German [15] demonstrated that the number of reviewers has an impact on the duration of the review process. Furthermore, Thongtanunam et al. [31] studied the characteristics of code review patches that attract more reviewers. They found that past review participation indicates that patches tend to have low participation, with patch description length and introduction of new features associated with likely low and delayed reviewer contribution. Maddila et al. [20] proposed a solution to predict the time to complete code review, with analyzing the code review stage and the actor that block the process. Baysal et al. [2] in an empirical study on Webkit and Google Blink revealed that both technical factors (e.g., code-related metrics) and nontechnical factors (e.g., human-related metrics) may impact the duration of the code review. To avoid losing time during the code review process, Fan et al. [7] and Islam et al. [14] leveraged machine learning to predict whether a code change will be merged in the base code, and the factors impacting this metric. To avoid wasting effort, Li et al. [19] and Khatoonabadi et al. [17] explored using machine learning duplicated and abandoned pull requests, respectively. Li et al. [19] revealed that duplicate pull requests are resource consuming, investigating the causes of duplication and the criteria developers use to select which pull requests to accept. Khatoonabadi et al. [17] found that abandoned pull requests often involve less experienced contributors, exhibit higher complexity, and undergo longer code review processes.

To improve the quality of code review, Hasan et al. [11] presented a comprehensive approach to automate the measurement of the quality of code reviews and address challenges faced by reviewers, suggesting that their approach can replace human evaluation. Dogˇan and Tu¨zu¨n [6] studied the code review smells. They found that the studied code reviews are impacted by at least one code review smell in 72.2% of cases.

The study by Wang et al. [34] and Wang, Bansal, and Nagappan [33] is the closest to our paper. They focused on characterizing code reviews with large review effort, requiring more than two iterations, to analyze change intents. Their findings highlighted that code reviews with specific intents, like refactoring, are more likely to require more large review effort. They also demonstrated the effectiveness of machine learning models in identifying code reviews that need large review effort, with models designed for specific intents outperforming those without considering change intent.

| Project | Nb of MRs | Min | Median | Mean | Max |
|---|---|---|---|---|---|
| Omnibus | 7217 | 0 | 14 | 95 | 5695 |
| Gitlab-runner | 4417 | 0 | 25 | 197 | 18492 |
| Inkskape | 6003 | 0 | 20 | 189 | 12727 |
| Gitaly | 6000 | 0 | 44 | 199 | 12685 |

TABLE I
AMOUNT OF CHANGES FOR ACCEPTING MERGE REQUESTS INSIGHTS OF THE STUDIED PROJECT

Our study differs in objectives. Firstly, they define large review efforts as review efforts measured as the number of iterations, not efforts to adjust code during the review. Secondly, we aim to explore factors that impact the amount of changes needed across various dimensions of metrics, such as history, experience, and collaboration, rather than specifically focusing on change intents. Lastly, they examined code reviews on Gerrit, while we focus on GitLab, two distinct code review platforms with different structures.

## III. DATA COLLECTION

In this study, we use the 4 open source projects shown in Table I from GitLab that have more than 3k MRs. The first project is Inkscape (Ink) [4] which is a vector image editor project. The second project is Omnibus-GitLab [5], which is a project to create full-stack platform-specific downloadable packages for GitLab. We also use GitLab-runner [6] that is used to run your CI/CD jobs for GitLab projects. Finally, we use Gitaly project [7] which is a service for handling all the git calls made by GitLab. To collect our dataset, we use the GitLab API to extract data related to MRs, commits, code changes, discussions, and review comments (notes). The data is available at: https://figshare.com/s/632a15054ffb881ae72d

## IV. DATA PROCESSING

The goal of this section is to discuss the steps that we followed to collect the metrics to train and test our models to explain the amount of changes for accepting MR. Below, we explain the broad outlines of our process, leaving the details to the research questions.

*1) Correlation & Redundancy Analysis:* Collinearity between features can affect the interpretability of machine learning models [29, 5]. Following prior work [16, 28], we apply correlation and redundancy analysis to reduce feature collinearity in each dataset. First, we use Spearman correlation with a 70% threshold [21, 18], applying dissimilarity dendrogram clustering to retain one feature from correlated groups (dissimilarity <0.3). For example, in the GitLab-runner project (Figure **??**), we retain the number of historical opened MRs, which captures the same information as its branch-specific counterpart. We also apply redundancy analysis using $R^2$ and exclude features explainable by others (threshold >90%) [30, 18].

*2) Effort Labeling and Noise Reduction:* We convert the effort data into "large" or "small" classes using median-based discretization to enable binary classification, as our goal is to explain the factors driving high or low review effort. To mitigate the impact of discretization noise on model performance and interpretation [23], we apply a noise reduction process based on maximizing AUC—a stable and reliable evaluation metric [12, 22, 9]. Noise is removed only from the training set, while the test and validation sets remain untouched. Median values are computed from the full dataset (see Table I).

## V. UNDERSTANDING AMOUNT OF CHANGES FOR ACCEPTING MRS

### RQ1: HOW MUCH CHANGES ARE REQUIRED TO MERGE A MR?

*Motivation*

This research question explores how much code is typically changed before an MR is accepted, and how this effort correlates with known review process metrics like review duration and reviewer participation [4, 31]. While prior studies have focused on timing and involvement to assess review quality [34, 20], the extent of changes required during review remains underexplored. Since more changes often imply more effort from both authors and reviewers, this metric offers a valuable lens for understanding review challenges. We analyze its distribution and assess whether it aligns with or differs from traditional review metrics.

*Approach*

In this study, we define the amount of changes to accept a MR as the total number of added and deleted lines (i.e., additions + deletions) in all commits made after the MR is created. We only consider commits that are committed after the MR creation, regardless of their authored date, since authored commits are not visible until they are committed. At the time of MR creation—our threshold—only committed changes are available and included in our analysis.

We analyze the distribution of the amount of changes across MRs and examine its correlation with key review metrics such as review duration, number of reviewers, committers, commenters, and total comments.

We use the Spearman correlation between different metrics [26], using the following interpretation:

- *Very High Correlation*, if 0.9¡ $|r— \leq 1$
- *High Correlation*, if 0.7¡ $|r— \leq 0.9$
- *Moderate Correlation*, if 0.5¡ $|r— \leq 0.7$
- *Low Correlation*, if 0.3¡ $|r— \leq 0.5$

[4]https://gitlab.com/inkscape/inkscape
[5]https://gitlab.com/gitlab-org/omnibus-gitlab
[6]https://gitlab.com/gitlab-org/gitlab-runner
[7]https://gitlab.com/gitlab-org/gitaly

| Metric | Description | Dimension | Hypothesis |
|---|---|---|---|
| is hashtag | true if the description or title contains a hashtag | Text | The text metrics may indicate a more informations about the merege request helping practitionners to identify mereg requests that requires large amount of changes. |
| is at tag | true if the description or title contains an at tag | | |
| description length | length of the description of the MR | | |
| title length | length of the title of the MR | | |
| **#commits** | total number of commits | Collaboration | A higher number of commits may indicate more incremental changes, whcih can increase the amounts of changes for accepting MRs |
| #unique committers | number of unique committers involved | | |
| mean time between commit | average time between commits | | |
| #discussions | number of comments left by authors | | |
| #minor author | number of authors that created the MR and participated in less than 5 percent of the changes of the historical MRs | Experience | The experience of authors can impact the quality of submitted code to the review, which can sometimes require larger amount of changes to be accepted |
| #major author | number of authors that created the MR and participated in more than 5 percent of the changes of the historical MRs | | |
| #approved fast integration | number of historical MRs where the number of changes is greater than 200 per hour | History | A higher number of fast integrations may indicate efficient collaboration and a streamlined code review process, potentially impacting amount of changes for accepting MRs. |
| #hist opened MR | number of historical MRs that are opened | | |
| #MR without discussion | the number of historical MRs without discussion | | |
| meantime approval normalized | mean of the time approval normalized by the number of changes of the historical MRs | | |
| mean nb discussions without bot normalized | mean of the number of discussions without bots normalized by the number of changes of historical MRs | | |
| number of self-approved MR | number of historical MRs where the creator is the integrator | | |
| #hist opened MR | number of historical MRs that are opened | | |
| **mean MR size hist** | mean size of the historical merged requests (initial size before creating the MR+ amount of changes for accepting MRs) | | |
| **mean amount of changes for accepting MRs hist** | mean amount of changes for historical merged requests | | |
| source branch workload | number of prior MRs on the same source branch | Branching | A higher number of priors MRs on the same source branch may indicate a more active development history, potentially impacting amount of changes for accepting MRs. |
| source branch in progress | number of prior MRs that are still opened on the same source branch | | |
| source branch delay | mean of the time of approval normalized by the number of changes of prior MRs on the same source branch | | |
| target branch workload | number of MRs with the same target branch | | |
| target branch in progress | number of prior MRs that are still opened with the same target branch | | |
| target branch delay | mean of the time of approval normalized by the number of changes of prior MRs on the same target branch | | |
| delta time | time between the creation of the project and the MR | Date | Longer delta time may indicate a more mature project, potentially impacting the development process and amount of changes for accepting MRs. |
| MR creation date | the date of creation of the MR | | |
| #file types | number of file types | Complexity | We suggest that having more complex code leads and more complex changes resulting in larger amount of changes for accepting MRs |
| #initial files | number of initial files included in a MR | | |
| initial mr size | initial size of the MR (added + deleted lines before the MR creation) | | |
| churn addition | number of initial lines added | | |
| churn deletion | number of initial lines deleted | | |
| change entropy | measure of the distribution of changes among initial files in prior MRs | | |

TABLE II
LIST OF COLLECTED METRICS

- *Negligible Correlation*, if 0.3¡ |*r*— ≤ 5

Note that we focus on the correlation between the amount of changes to accept a MR and two key review metrics: code review duration and number of reviewers, as these have been widely studied in prior work. A strong correlation may suggest that findings related to review duration and reviewer count could also apply to the amount of changes. Otherwise, if the correlation is weak, it implies that prior conclusions may not generalize, and a separate analysis of the amount of changes is necessary.

*Results*

**66%, 66%, 68% and 73% of the analyzed MRs require an amount of changes during the code review process for Omnibus, GitLab Runner, Gitaly, and Inkscape, respectively.**. The density distribution in Figure 1 highlights a concentration of MRs with an amount of changes in the 1–200 range, with an average of 38% across all projects. Among the MRs that require changes, up to 28% involve at least 200 lines of code being modified during the review process. In addition, we observe that certain MRs exhibit more than 1,000 changed lines across all projects. For instance, within the Omnibus (OB) project, out of 7,217 MRs, 4,739 require changes after submission. Among these, 2,861 involve a change size between 1–200 lines, while 1,878 require at least 200 lines of changes. Notably, 76 MRs within this group show an amount of changes exceeding 1,000 lines—reaching up to 5,695, as shown in Table 1. These findings underscore the importance of managing the amount of changes required to accept a MR, as it not only increases the size and complexity of a MR but also raises the risk of errors, redundancy, and communication challenges. It can further strain contributors, especially when working under tight deadlines.

**The number of additions and deletions are highly correlated across all studied projects, with Spearman correlation coefficients exceeding 0.9.** Specifically, we observe values of 0.97, 0.96, 0.96, and 0.92 for the Omnibus, GitLab Runner, Inkscape, and Gitaly projects, respectively—indicating a strong association between added and deleted lines of code. These results suggest that most changes within MRs involve modifications to existing code, rather than purely new additions or removals. This dynamic emphasizes that the amount of changes to accept a MR may affect code stability, which is often considered an indicator of overall software product stability [27].

**We observe low and negligible correlations between the amount of changes to accept a MR and both code review duration and the number of people involved in the MR.** The Spearman correlation values between code review duration and the amount of changes are 44%, 42%, 47%, and 48% for

the Omnibus, GitLab Runner, Inkscape, and Gitaly projects, respectively. This suggests that the time taken to complete a code review is not strongly related to the size of the code changes made during the process.

In contrast, the correlation between the amount of changes and the number of comments is notably higher—71%, 67%, 64%, and 67% for Omnibus, GitLab Runner, Inkscape, and Gitaly, respectively—indicating that developers often need to review, comment on, and revise code within the same limited time frame.

The Spearman correlation between the number of contributors and the amount of changes is low—20%, 19%, 29%, and 19% for the four projects—suggesting no meaningful relationship between how many people are involved and the size of the changes required. For example, in the Inkscape project, a MR with 1,200 lines of changes involves only three contributors, while a change-free MR involves ten. Similarly, in GitLab Runner, a MR with about 7,500 lines of changes involves six reviewers, whereas one with 2,500 changes involves 13 contributors. These observations reinforce the conclusion that the amount of changes to accept a MR is largely independent of the number of collaborators involved.

## VI. Predicting amount of changes for accepting MRs

### RQ2: How effective are machine learning models at classifying the amount of changes in MRs?

*Motivation*

In addressing this RQ, our goal is to investigate the effectiveness of different machine learning models in classifying MRs according to their amount of changes before acceptance, with the goal of identifying the most accurate classifier for subsequent analysis. Understanding the amount of changes required for a given MR provides invaluable insight into the code review process, allowing teams to optimize both time and human effort. Similar to previous studies, we use machine learning to classify MRs based on their change size. This task requires a robust ML model with high performance and reliability. Therefore, this research investigates the effectiveness of ML models in classifying the amount of changes of MRs.

*Approach*

To evaluate the effectiveness of machine learning (ML) models in classifying the the amount of changes of MRs, we conduct a comparative analysis involving four distinct classifiers: K nearest neighbor **(KNN)**, Classification and Regression Trees **(CART)**, Random Forest **(RF)**, and Logistic Regression **(LR)**, which have been used in previous studies [24, 23, 16]. Our approach is structured as follows:

- **Model Training:** To explain the amount of changes for accepting MRs, we evaluate four different classifiers K nearest neighbour, random forest, decision tree, and logistic regression. We use the 100 out-of-sample bootstrap technique similar to prior studies [24, 29] for obtaining statistically robust results and conclusions. We generate the samples for both training and testing sets in each iteration, while we use the validation set for all the bootstraps samples. We ensure that we have 10 data points for each feature to train the model (e.g. if we have 20 features so we need at least 200 MRs).
- **Performance Evaluation:** To evaluate our model, we divided the data set into training and test sets, with 80% of the data in the training set and 20% of the data in the validation set. The validation set consists of the most recent MRs in terms of creation date. For evaluation metrics, we use the commonly used metrics in the literature [23, 8, 30, 36], which are Accuracy (ACC), Precision (PRC), Recall (RCL), Brier Score (BS), Area Under Curve (AUC), F-measure (FM), and Mathew's Correlation Coefficient (MCC). We generate 100 bootstrap samples for each metric, resulting in 100 values for each sample for the and 100 values for each sample for the validation.
- **Classifier Ranking:** We rank all ML regressors performance for both test and validation using Scott-Knott Effect Size Difference (ESD) Test [8], which clusters distributions into statistically distinct ranks.
- **Statistical Difference:** To identify subtle differences in performance between two classifiers when their performance is closely matched, we use the Anova test with a significance level of p-value = 0.05. This statistical test compares the distributions of a given metric. If the resulting p-value is less than 0.05, we interpret it as indicating a significant difference between the distributions. Conversely, if the p-value is greater than 0.05, we conclude that there is no statistically significant difference between them.

Next, we assess the importance of each metric dimension in the best-performing model by iteratively removing one dimension at a time and retraining the model on 100 bootstrap samples. We compare the median performance (across bootstraps) of each reduced model to that of the full-feature model to determine the impact of each dimension. Statistical significance is evaluated using ANOVA to identify meaningful performance differences, providing insight into each dimension's contribution to classification accuracy.

- **Model Training:** By iteratively removing one dimension at a time, we train the model on 100 bootstrap samples. The comparative analysis of the model performance, taking into account the removal of metric dimensions, allows the ranking of the importance of the dimensions.
- **Impact:** we consider the impact of removing one dimension on the baseline model (model with all features) as the median of 100 bootstraps of the model without dimension on the median of the 100 bootstraps of the baseline model.
- **Statistical Difference:** similarly to the previous step, we use ANOVA test to measure the statistical distribution significance between the performance metrics, providing

[8] https://cran.r-project.org/web/packages/ScottKnottESD/

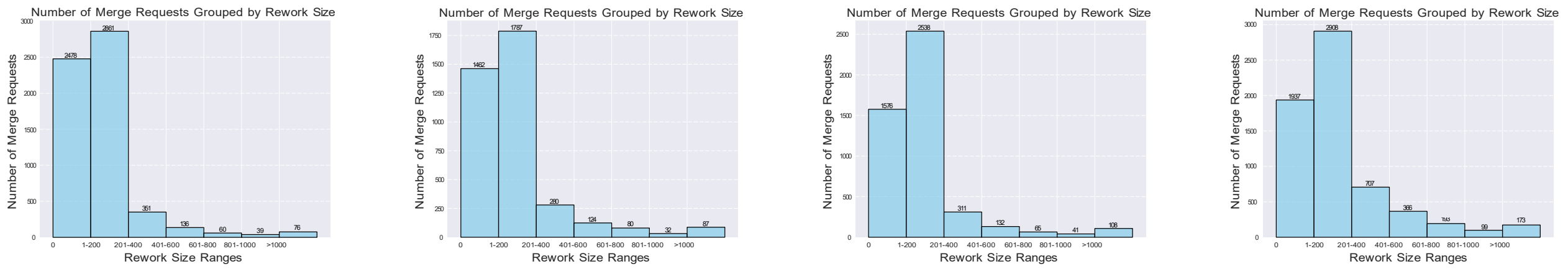

(a) Omnibus (b) GitLab-runner (c) Inkscape (d) Gitaly

Fig. 1. Distribution of MRs by the amount of changes required for acceptance in four projects.

valuable insight into the contribution of each dimension in the classification accuracy.

*Results*

**Across all projects studied, the Random Forest (RF) classifier shows the best performance on 67% of the measured metrics when evaluated on test datasets.** As shown in the Figure 2, particularly on the GR, GT, and Ink projects, random forest outperforms other classifiers in all metrics except recall where Classification and Regression Trees (CART) occurs as the top-1 classifier. However, for the OB project, RF outperforms other classifiers only in precision, with CART outperforming in all other metrics. Interestingly, when RF doesn't take the top 1 position, its performance metrics closely align with CART, suggesting a subtle difference between these two classifiers.

**The good performance of our models are consistent on the validation datasets, whereas RF achieves the best performance in 96% of the measured metrics**, as shown in Figure 3. This consistency confirms the reliability of our model in classifying unseen MRs based on their amount of changes to be accepted. In our comparative analysis of classifiers, RF classifier achieves the first rank with an AUC range from 0.84 and 0.88, except for the recall metric for the GT and OB projects, where CART performs best. However, in both case studies, RF outperforms the other classifiers in all other metrics. To evaluate the significance of performance differences between RF and other classifiers, we perform ANOVA test. The results presented in table III indicate a significant difference in the distribution of 100 bootstrapped performances between RF and other classifiers for all case studies, underscoring that RF consistently achieves significantly better results compared to the other classifiers.

**Our analysis shows that excluding any metric dimension results in a significant decrease in performance across all projects studied, suggesting that all the used dimensions are relevant in the classification of MRs based on their amount of changes.** Figure 4 illustrates that using the random forest classifier identified as the best model for amount of changes classification, the removal of various metric dimensions results in a consistent decrease in performance across all studied projects. When examining the impact, as shown in Figure 5, we find that removing any metric dimension negatively impacts all the metrics, except for the GR project. In the case of the GR project, only the removal of the complexity dimension results in a negative impact on the model, whereas the removal of other dimensions yields a positive impact.

| *Dataset* | *Classifier* | *1-MSE* | *Precision* | *Recall* | *AUC* | *MCC* | *ACC* | *F1* |
|---|---|---|---|---|---|---|---|---|
| OB | CART | **1.2e-20** | **1.4e-04** | **1.3e-45** | **1.0e-20** | **5.4e-20** | **1.2e-20** | **4.8e-25** |
| | LR | **5.1e-78** | **1.2e-60** | **4.1e-64** | **1.2e-77** | **1.7e-77** | **5.1e-78** | **3.0e-89** |
| | KNN | **1.1e-193** | **2.6e-173** | **2.4e-138** | **4.2e-194** | **3.4e-194** | **1.1e-193** | **1.2e-178** |
| GR | CART | **5.6e-20** | **8.7e-38** | **5.3e-05** | **1.4e-18** | **2.2e-20** | **5.6e-20** | **2.8e-15** |
| | LR | **4.9e-92** | **8.6e-60** | **2.4e-87** | **1.9e-94** | **1.1e-90** | **4.9e-92** | **1.8e-91** |
| | KNN | **1.2e-157** | **7.9e-146** | **6.2e-116** | **2.0e-157** | **4.2e-158** | **1.2e-157** | **1.1e-149** |
| GT | CART | **3.1e-23** | **7.3e-45** | **5.9e-04** | **2.7e-25** | **2.7e-24** | **3.1e-23** | **4.4e-19** |
| | LR | **2.4e-110** | **1.4e-56** | **3.4e-114** | **6.4e-110** | **5.0e-109** | **2.4e-110** | **2.1e-108** |
| | KNN | **1.5e-172** | **1.9e-155** | **1.6e-117** | **1.5e-173** | **4.1e-173** | **1.5e-172** | **2.6e-162** |
| Ink | CART | **3.1e-30** | **1.4e-62** | **5.3e-07** | **3.0e-24** | **5.7e-31** | **3.1e-30** | **1.6e-23** |
| | LR | **1.3e-149** | **5.6e-97** | **3.2e-142** | **1.6e-153** | **2.4e-145** | **1.3e-149** | **2.0e-150** |
| | KNN | **7.9e-183** | **1.8e-163** | **2.3e-124** | **3.3e-180** | **6.4e-183** | **7.9e-183** | **1.7e-169** |

TABLE III
ANOVA TEST OF RF WITH OTHER CLASSIFIERS (P-VALUE)

## RQ3: WHAT ARE THE MOST IMPORTANT FEATURES THAT INFLUENCE THE AMOUNT OF CHANGES REQUIRED TO ACCEPT A MR?

*Motivation*

In addressing this research question, our goal is to examine the factors that impact the amount of changes required to accept a MR. Analyzing the metrics that contribute to larger change sizes provides practitioners with valuable insights into what drives significant code modifications during the review process. By identifying these influential factors, teams can establish best practices to reduce human effort, improve code quality, and ultimately minimize the extent of required changes. This section explores the key features impacting the amount of changes in each dataset and offers an interpretation of how each metric contributes to this outcome.

*Approach*

In this step, we use Random Forest Classifier (RF), that shows the best performance in RQ2, to study the factors that impact the code review metrics. To do so, we follow these steps:

- **Perform permutation analysis:** this technique is used to assess the statistical significance of feature importance by comparing actual model performance with random permutations in the dataset independent metrics, providing a robust method for identifying key factors that explain the dependent variable. We apply permutation analysis on each of 100 generated bootstraps using validation dataset to assess the feature importance.

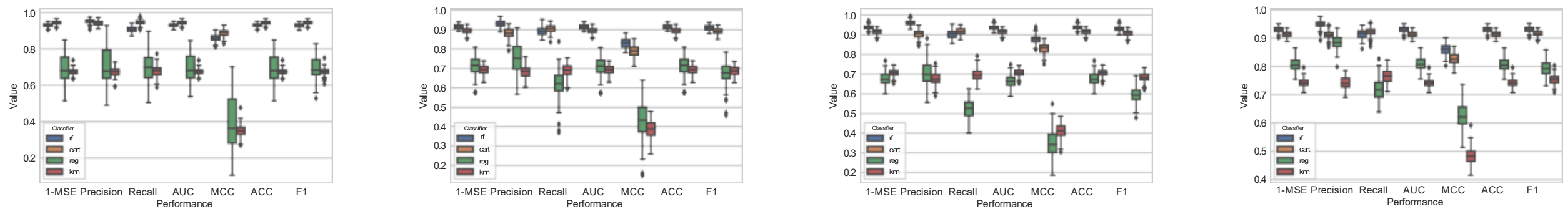

(a) OB (b) GR (c) Ink (d) Gt

Fig. 2. Comparative multi box plots for performance distribution on test datasets

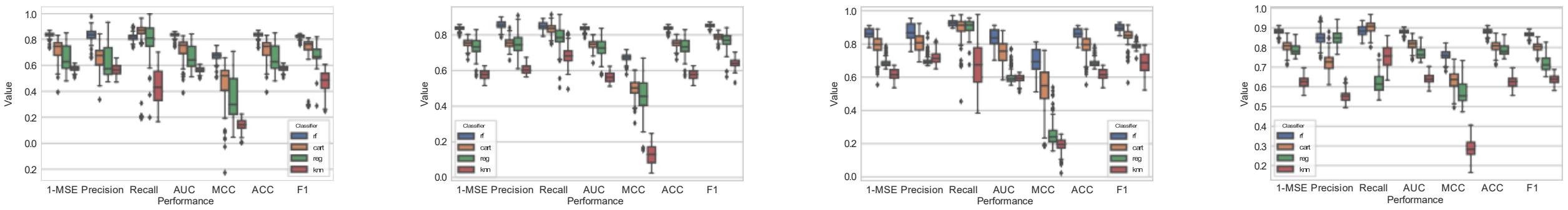

(a) OB (b) GR (c) Ink (d) GT

Fig. 3. Comparative multi box plots for performance distribution on validation datasets

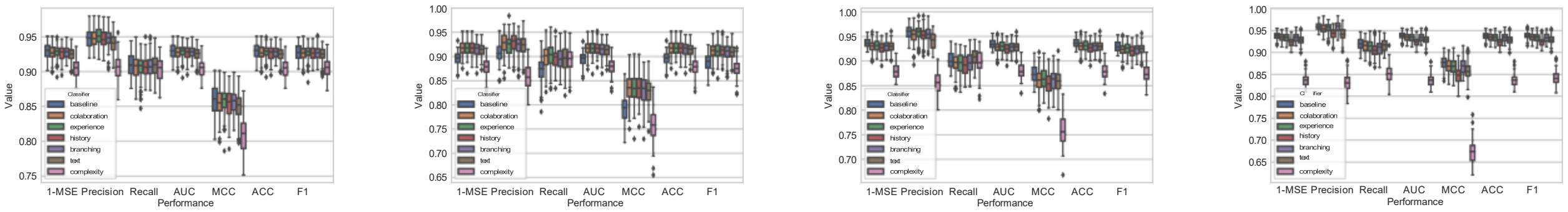

(a) OB (b) GR (c) Ink (d) GT

Fig. 4. Comparative multi box plots for performance distribution when removing metric dimensions

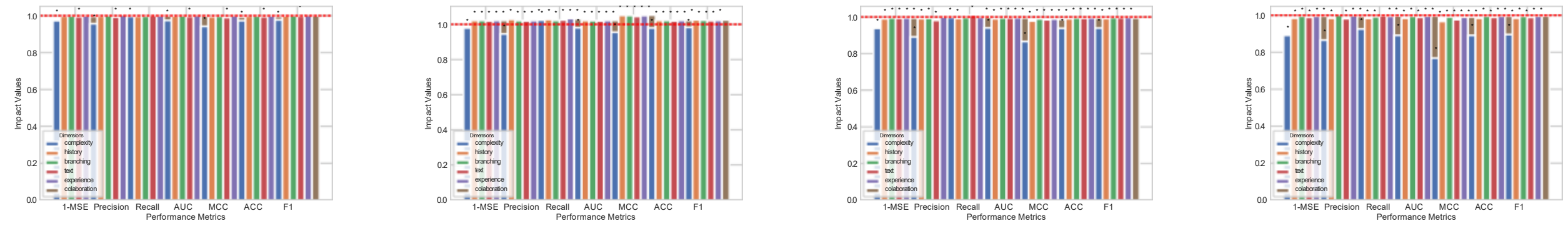

(a) OB (b) GR (c) Ink (d) GT

Fig. 5. Comparative multi bar chart for performance distribution when removing metric dimensions
- The bars marked with an asterisk (*) indicate significant impact using the Anova test.
- The horizontal line helps to differentiate the positive and negative impacts.

- **Rank the features across all 100 bootstrap samples:** Similarly to the performance ranking, we rank features by their importance while accounting for variability of ranks across different samples using Scott-Knott Effect Size Difference (ESD) Test.
- **Analyze associations between metrics and the amount of changes:** Using Accumulated Local Effects (ALE) plots, we explore the relationships between independent variables and the amount of changes required to accept a MR, in order to understand their impact on the classification outcome. Following the approach of Khatoonabadi et al. [17], we exclude feature values exceeding the 99th percentile in each project to reduce the influence of outliers. A negative impact suggests a higher likelihood of a small amount of changes, while a positive impact indicates a higher probability of a large amount of changes for the given MR.

*Results*

**Our analysis shows that complexity metrics are the two most important factors influencing the amount of changes of a MR**. As illustrated in Figure 6, for all the studied projects, the number of initial files, followed by the initial size of the MR, rank as the first and second most important features, respectively. As shown in Figure 7, the increasing number of initial files implies a larger amount of changes. We suggest that a higher number of initial files in a MR indicates a more complex change during code review, especially if those files are interdependent. Interdependency in this context

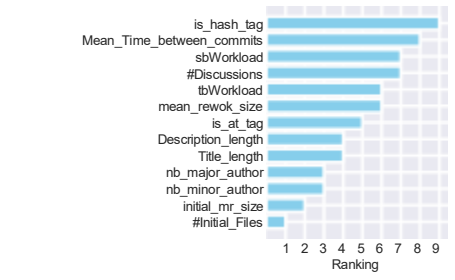

(a) OB

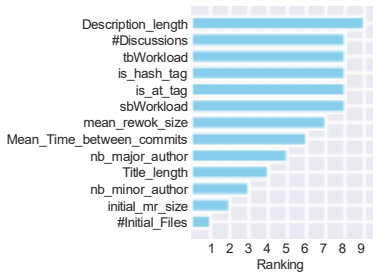

(b) GR

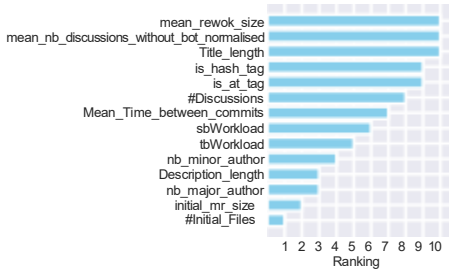

(c) Ink

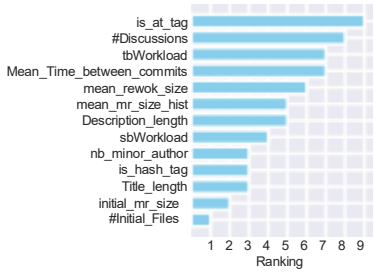

(d) GT

Fig. 6. SK-ESD Feature Importance ranks

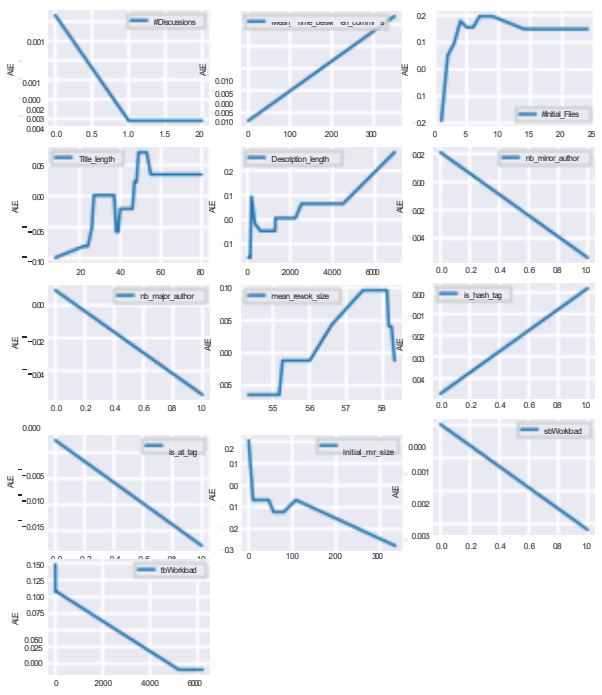

(a) OB

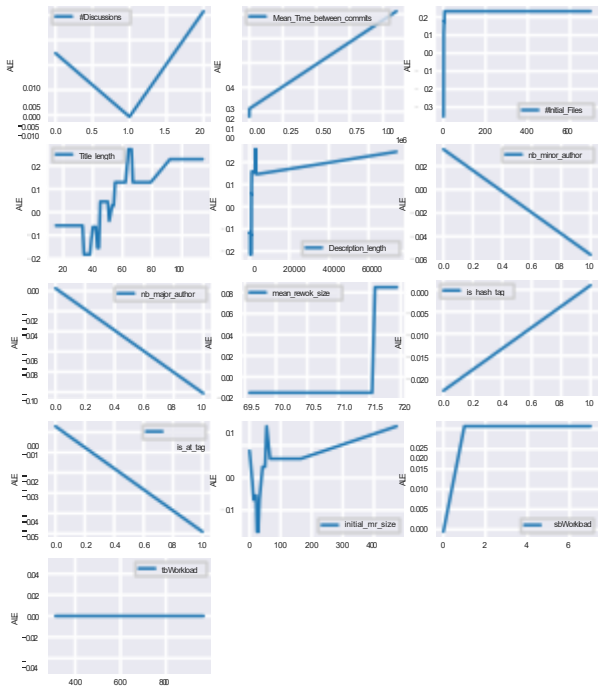

(b) GR

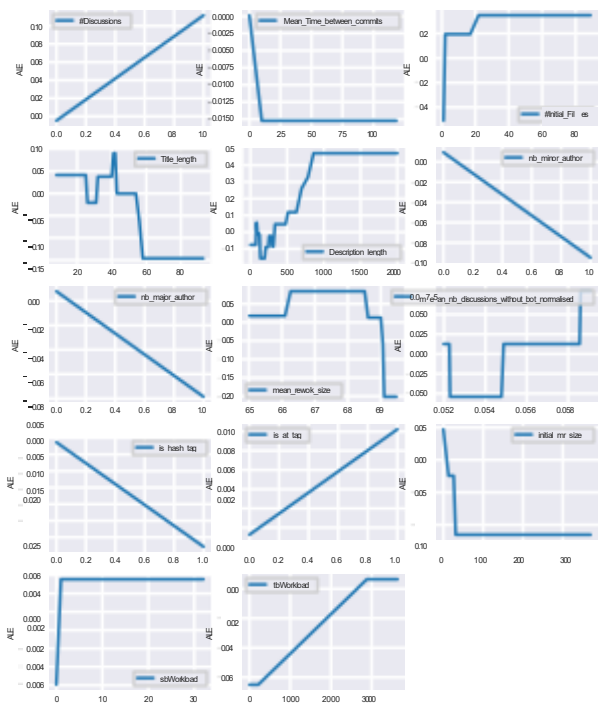

(c) Ink

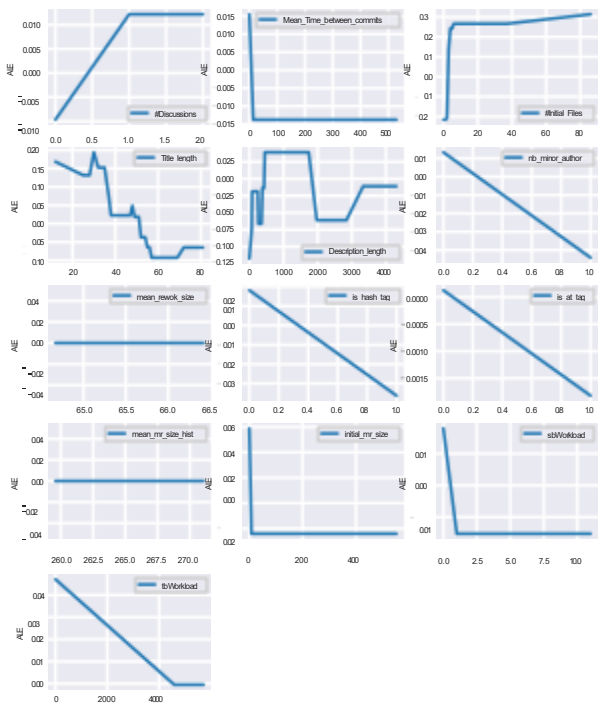

(d) GT

Fig. 7. Feature Importance Impact On the Classification Model

indicates that a change in one file requires updates in all related files, thereby increasing the amount of changes. For example, changing an attribute in a configuration file requires updates in all associated files, increasing the number of lines of code to change. At the same time, except the GR project, we observe that a larger initial size of a MR tends to correspond to a smaller amount of changes required during the code review process. This pattern suggests that in some cases, MRs with large initial sizes indicate that necessary changes are being made early on, before the MR is created. Examples of such purposes include adding documentation, changing branches, or improving modularity by splitting existing code into multiple files. These actions involve significant amount of the lines of code modified before the MR is created, but typically do not require code changes during the review. Conversely, smaller initial MRs may indicate the addition of new configurations or features. Such an introduction often requires extensive code changes for integration into the main codebase, resulting in the need for more review and larger amount of changes.

**Author experience metrics also have a significant impact on amount of changes required to accept a MR, consistently ranking in the top-3 or top-4 in 3/4 of studied projects.** Figure 6 shows that the number of major authors ranks as the third most important feature for OB and Ink projects, and fifth for GR projects. As shown in Figure 7, the impact of this metric is negative for all projects, suggesting that having more major authors, who typically have experience with the project, is associated with smaller amount of changes. We interpret this by experienced authors contribute code that requires fewer adjustments or revisions during the review process. In the other hand, we observe that the number of minor authors ranks third in importance for GR, GT, and OB projects, and fourth for the Ink project. In all the studied projects, the impact of the number of minor authors on the classification of amount of changes is consistently negative. This implies that the presence of minor authors tends to result in smaller amount of changes. We suggest that collaboration between major and minor authors may represent effective mentorship and knowledge transfer within the team. Experienced authors guiding less experienced contributors can result in code that require less changes, highlighting the positive impact of such collaborative dynamics on code review.

**Text-related metrics consistently rank among the top five most important features across all projects.** As shown in Figure 6, description length ranks third in Ink, fourth in OB, and fifth in GT. Figure 7 shows its consistently positive impact, suggesting that longer descriptions are associated with larger code changes. This likely reflects submissions requiring major revisions—such as new features or services—before merging. For instance, in GT, a MR introducing distributed reads had a 3997-character description and 2995 lines of code changed[9], while another with no description simply fixed a typo with zero changes[10].

Title length is also important, ranking in the top five for GR, OB, and GT. However, its impact varies—positive in half the projects and negative in the rest—suggesting that titling practices differ by context.

Tag-related metrics also appear in the top five for OB and GT. The presence of an "@ tag" generally correlates with fewer changes in 3 out of 4 projects, possibly because tagged individuals focus the review on their area of responsibility. In contrast, hashtags show project-specific effects. Their meaning varies—from technical tags like bugfix or feature to communication tags like teamA—which may explain their inconsistent impact on change size.

[9]https://gitlab.com/gitlab-org/gitaly/-/merge_requests/2738
[10]https://gitlab.com/gitlab-org/gitaly/-/merge_requests/4422

**Historical metrics also show an impact on amount of changes in two projects.** Figure 6 shows that for the Ink project, the workload (number of MRs) on the source and target branches impacts the amount of changes. The positive impact of these metrics suggests that an increased workload can potentially reduce code quality by having more to-do tasks, which results in large amount of changes. For the GT project, the average of the amount of changes of historical MRs ranks in the top five. We suggest that the skill level of the individuals involved in the project requires more changes to produce the required quality of code. In addition, we suggest that this observation may be related to the nature of the project, which involves substantial MRs such as new feature additions or updates, that can require more changes.

## VII. Threats to validity

**Internal validity threats:** These may stem from implementation errors and model choices. We reduced such risks by using established metric definitions, widely adopted classifiers, 100-bootstrapping for reliability, and well-known Python packages (e.g., Scikit-learn). Nonetheless, unnoticed errors and context-specific result variations remain possible.

**External validity threats:** These concern the generalizability of our findings. Code review metrics based on the amount of changes can vary across project types and team structures. To mitigate this, we analyzed over 23.6K reviews across multiple projects and dimensions. However, expanding the feature set or studying other platforms (e.g., Gerrit) could improve coverage.

## VIII. Conclusion

In conclusion, code review is a critical aspect of ensuring code quality and knowledge transfer in software development. Estimating the effort invested in this process is essential for optimizing development practices. The introduction of amount of changes for accepting MRs, which measures the total number of lines of code changed after the creation of the MR, provides a quantifiable metric for understanding developers' efforts in addressing review comments. In this study, we use four open source projects on GitLab to investigate the factors that impact amount of changes for accepting MRs in MRs.

Our results show that a significant amount, up to 73%, of MRs are amount of changes for accepting MRed before integration. Interestingly, the amount of changes for accepting MRs is not correlated with code review duration or the number of people involved, suggesting that developers expend more effort with the same resources. Furthermore, we infer that the factors impacting time and people are different from those impacting amount of changes for accepting MRs. Using machine learning, we compared classification models and achieved an AUC of up to 0.88 in explaining the amount of changes for accepting MRs. The analysis highlights the consistent impact of complexity, experience, and text features across projects, with historical factors also sometimes influencing current MRs. For future research, extending this study to an industrial context is critical. In addition, exploring additional metrics to explain the code review process will contribute to in-depth understanding.

## References

[1] Gabriele Bavota and Barbara Russo. "Four eyes are better than two: On the impact of code reviews on software quality". In: *2015 IEEE International Conference on Software Maintenance and Evolution (ICSME)*. IEEE. 2015, pp. 81–90.

[2] Olga Baysal et al. "Investigating technical and non-technical factors influencing modern code review". In: *Empirical Software Engineering* 21 (2016), pp. 932–959.

[3] Amiangshu Bosu and Jeffrey C Carver. "Impact of developer reputation on code review outcomes in oss projects: An empirical investigation". In: *Proceedings of the 8th ACM/IEEE international symposium on empirical software engineering and measurement*. 2014, pp. 1–10.

[4] Moataz Chouchen et al. "Learning to Predict Code Review Completion Time In Modern Code Review". In: *Empirical Software Engineering* 28.4 (2023), p. 82.

[5] Ju¨rgen Cito et al. "Explaining mispredictions of machine learning models using rule induction". In: *Proceedings of the 29th ACM Joint Meeting on European Software Engineering Conference and Symposium on the Foundations of Software Engineering*. 2021, pp. 716–727.

[6] Emre Dogˇan and Eray Tu¨zu¨n. "Towards a taxonomy of code review smells". In: *Information and Software Technology* 142 (2022), p. 106737.

[7] Yuanrui Fan et al. "Early prediction of merged code changes to prioritize reviewing tasks". In: *Empirical Software Engineering* 23 (2018), pp. 3346–3393.

[8] Lina Gong et al. "Revisiting the impact of dependency network metrics on software defect prediction". In: *IEEE Transactions on Software Engineering* 48.12 (2021), pp. 5030–5049.

[9] Chongomweru Halimu, Asem Kasem, and SH Shah Newaz. "Empirical comparison of area under ROC curve (AUC) and Mathew correlation coefficient (MCC) for evaluating machine learning algorithms on imbalanced datasets for binary classification". In: *Proceedings of the 3rd international conference on machine learning and soft computing*. 2019, pp. 1–6.

[10] Kazi Amit Hasan et al. "Understanding the Time to First Response In GitHub Pull Requests". In: *arXiv preprint arXiv:2304.08426* (2023).

[11] Masum Hasan et al. "Using a balanced scorecard to identify opportunities to improve code review effectiveness: An industrial experience report". In: *Empirical Software Engineering* 26 (2021), pp. 1–34.

[12] Jin Huang and Charles X Ling. "Using AUC and accuracy in evaluating learning algorithms". In: *IEEE Transactions on knowledge and Data Engineering* 17.3 (2005), pp. 299–310.

[13] Yuan Huang et al. "Would the patch be quickly merged?" In: *Blockchain and Trustworthy Systems: First International Conference, BlockSys 2019, Guangzhou, China, December 7–8, 2019, Proceedings 1*. Springer. 2020, pp. 461–475.

[14] Khairul Islam et al. "Early prediction for merged vs abandoned code changes in modern code reviews". In: *Information and Software Technology* 142 (2022), p. 106756.

[15] Yujuan Jiang, Bram Adams, and Daniel M German. "Will my patch make it? and how fast? case study on the linux kernel". In: *2013 10th Working Conference on Mining Software Repositories (MSR)*. IEEE. 2013, pp. 101–110.

[16] J Jiarpakdee, C Tantithamthavorn, and AE Hassan. "The Impact of Correlated Metrics on the Interpretation of Defect Prediction Models". In: *IEEE Trans. Software Eng. early access* 10 (2019).

[17] SayedHassan Khatoonabadi et al. "On Wasted Contributions: Understanding the Dynamics of Contributor-Abandoned Pull Requests–A Mixed-Methods Study of 10 Large Open-Source Projects". In: *ACM Transactions on Software Engineering and Methodology* 32.1 (2023), pp. 1–39.

[18] Daniel Lee et al. "An empirical study of the characteristics of popular Minecraft mods". In: *Empirical Software Engineering* 25 (2020), pp. 3396–3429.

[19] Zhixing Li et al. "Redundancy, context, and preference: An empirical study of duplicate pull requests in OSS projects". In: *IEEE Transactions on Software Engineering* 48.4 (2020), pp. 1309–1335.

[20] Chandra Maddila et al. "Nudge: Accelerating Overdue Pull Requests toward Completion". In: *ACM Transactions on Software Engineering and Methodology* 32.2 (2023), pp. 1–30.

[21] Shane McIntosh et al. "The impact of code review coverage and code review participation on software quality: A case study of the qt, vtk, and itk projects". In: *Proceedings of the 11th working conference on mining software repositories*. 2014, pp. 192–201.

[22] Douglas Mossman. "Assessing predictions of violence: being accurate about accuracy." In: *Journal of consulting and clinical psychology* 62.4 (1994), p. 783.

[23] Gopi Krishnan Rajbahadur et al. "Impact of discretization noise of the dependent variable on machine learning classifiers in software engineering". In: *IEEE Transactions on Software Engineering* 47.7 (2019), pp. 1414–1430.

[24] Gopi Krishnan Rajbahadur et al. "The impact of using regression models to build defect classifiers". In: *2017 IEEE/ACM 14th International Conference on Mining Software Repositories (MSR)*. IEEE. 2017, pp. 135–145.

[25] Nishrith Saini and Ricardo Britto. "Using machine intelligence to prioritise code review requests". In: *2021 IEEE/ACM 43rd International Conference on Software Engineering: Software Engineering in Practice (ICSE-SEIP)*. IEEE. 2021, pp. 11–20.

[26] Charles Spearman. "The proof and measurement of association between two things." In: (1961).

[27] Miroslaw Staron et al. "Measuring and visualizing code stability–a case study at three companies". In: *2013 Joint Conference of the 23rd International Workshop on Software Measurement and the 8th International Conference on Software Process and Product Measurement*. IEEE. 2013, pp. 191–200.

[28] Chakkrit Tantithamthavorn and Ahmed E Hassan. "An experience report on defect modelling in practice: Pitfalls and challenges". In: *Proceedings of the 40th International conference on software engineering: Software engineering in practice*. 2018, pp. 286–295.

[29] Chakkrit Tantithamthavorn et al. "An empirical comparison of model validation techniques for defect prediction models". In: *IEEE Transactions on Software Engineering* 43.1 (2016), pp. 1–18.

[30] Chakkrit Tantithamthavorn et al. "The impact of automated parameter optimization on defect prediction models". In: *IEEE Transactions on Software Engineering* 45.7 (2018), pp. 683–711.

[31] Patanamon Thongtanunam et al. "Review participation in modern code review: An empirical study of the android, Qt, and OpenStack projects". In: *Empirical Software Engineering* 22 (2017), pp. 768–817.

[32] Rosalia Tufano et al. "Towards automating code review activities". In: *2021 IEEE/ACM 43rd International Conference on Software Engineering (ICSE)*. IEEE. 2021, pp. 163–174.

[33] Song Wang, Chetan Bansal, and Nachiappan Nagappan. "Large-scale intent analysis for identifying large-review-effort code changes". In: *Information and Software Technology* 130 (2021), p. 106408.

[34] Song Wang et al. "Leveraging change intents for characterizing and identifying large-review-effort changes". In: *Proceedings of the Fifteenth International Conference on Predictive Models and Data Analytics in Software Engineering*. 2019, pp. 46–55.

[35] Xunhui Zhang et al. "Pull request latency explained: An empirical overview". In: *Empirical Software Engineering* 27.6 (2022), p. 126.

[36] Thomas Zimmermann and Nachiappan Nagappan. "Predicting defects using network analysis on dependency graphs". In: *Proceedings of the 30th international conference on Software engineering*. 2008, pp. 531–540.